# A review of mechanistic learning in mathematical oncology

John Metzcar[1,2], Catherine R. Jutzeler[3,4], Paul Macklin[1], Alvaro Köhn-Luque[6,7], Sarah C. Brüningk[3,4*]

- * Corresponding author: sarah.brueningk@hest.ethz.ch
- [1] Intelligent Systems Engineering, Luddy School of Informatics, Computing, and Engineering, Bloomington, Indiana, USA
- [2] Informatics, Luddy School of Informatics, Computing, and Engineering, Bloomington, Indiana, USA
- [3] ETH Zürich, Department of Health Sciences and Technology (D-HEST), Zürich, Switzerland
- [4] Schulthess Klinik, Zürich, Switzerland
- [5] Swiss Institute of Bioinformatics (SIB), Lausanne, Switzerland
- [6] Oslo Centre for Biostatistics and Epidemiology, Faculty of Medicine, University of Oslo, Oslo, Norway
- [7] Oslo Centre for Biostatistics and Epidemiology, Research Support Services, Oslo University Hospital, Oslo, Norway

## Abstract

Mechanistic learning, the synergistic combination of knowledge-driven and data-driven modeling, is an emerging field. In particular, in mathematical oncology, the application of mathematical modeling to cancer biology and oncology, the use of mechanistic learning is growing. This review aims to capture the current state of the field and provide a perspective on how mechanistic learning may further progress in mathematical oncology. We highlight the synergistic potential of knowledge-driven mechanistic mathematical modeling and data-driven modeling, such as machine and deep learning. We point out similarities and differences regarding model complexity, data requirements, outputs generated, and interpretability of the algorithms and their results. Then, organizing combinations of knowledge- and data-driven modeling into four categories (sequential, parallel, intrinsic, and extrinsic mechanistic learning), we summarize a variety of approaches at the interface between purely data- and knowledge-driven models. Using examples predominantly from oncology, we discuss a range of techniques including physics-informed neural networks, surrogate model learning, and digital twins. We see that mechanistic learning, with its intentional leveraging of the strengths of both knowledge and data-driven modeling, can greatly impact the complex problems of oncology. Given the increasing ubiquity and impact of machine learning, it is critical to incorporate it into the study of mathematical oncology with mechanistic learning providing a path to that end. As the field of mechanistic learning advances, we aim for this review and proposed categorization framework to foster additional collaboration between the data- and knowledge-driven modeling fields. Further collaboration will help address difficult issues in oncology such as limited data availability, requirements of model transparency, and complex input data.

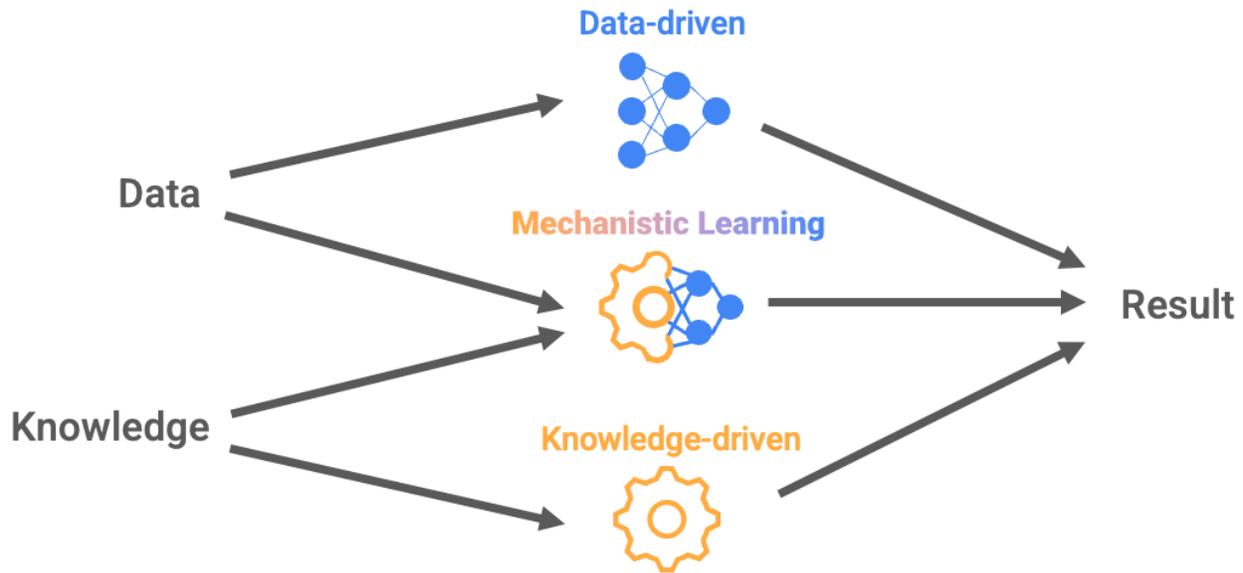

***Graphical Abstract:*** *Data and knowledge both drive the progress of research and are the cornerstones of modeling. Depending on the emphasis, both **data-driven** (exemplified by machine and deep learning) and **knowledge-driven** (exemplified by mechanistic mathematical modeling) models generate novel results and insights. **Mechanistic learning** describes approaches that employ both data and knowledge in a complementary and balanced way.*

# 1. Introduction

An increasing understanding of cancer evolution and progression along with growing multi-scale biomedical datasets, ranging from molecular to population level, is driving the research field of mathematical oncology[1]. Mathematical oncology aims to bridge the gaps between medicine, biology, mathematics, and computer science to advance cancer research and clinical care. Both data and understanding of cancer biology contribute to this aim.

Data science may be defined as "*a set of fundamental principles that support and guide the principled extraction of information and knowledge from data*"[2]. Here, problem-solving is approached from the perspective of a learning process accomplished through observing diverse examples[3]. Relationships between various types of input data (e.g., omics and imaging) and outcomes (e.g., overall survival) are abstracted where a mechanistic understanding of a relationship is missing or otherwise not accounted for.

An alternative is to formulate a specific guess on how relevant variables interact between input and output through the formulation of a mathematical model. Bender defines a mathematical model as an "*abstract, simplified, mathematical construct related to part of reality and created for a particular purpose*"[4]. Here the formulation of deliberate approximations of reality through equations or rules is key[5]. In turn, the quality and limits of this approximation are validated with data. Independent of the use of a data science or a mathematical modeling formulation, "data" and "knowledge" are indispensable. The emphasis on data and knowledge may vary leading to the terminology of "data-driven" and "knowledge-driven" modeling[6]. The fluid boundaries between these concepts motivate their combination.

The evolving field of mechanistic learning[7] aims to describe synergistic combinations of classical mathematical modeling and machine learning[8,9]. In this review, we provide an overview of the key aspects of these approaches, explain possible ways of combining them, present a selection of examples, and discuss how mechanistic learning can thrive in mathematical oncology. In doing so, we aim to draw awareness to similarities and synergies between knowledge- and data-driven modeling, noting that this combination could help push mathematical oncology into the clinic[10] as reliable, data-supported, and explainable models in the context of oncology.

## 2. Contrasting "knowledge-driven" and "data-driven" modeling"

As per definition, data- and knowledge-driven modeling are complementary perspectives for approaching research questions. Here, we address similarities and differences to understand synergies at the interface of these fluid concepts.

### 2.1. Knowledge-driven modeling approximates biomedical understanding

According to Rockne et al.[1], the goal of knowledge-driven modeling is to describe the behavior of complex systems based on an understanding of the underlying mechanisms rooted in fundamental principles of biology, chemistry, and physics. While the formulation of the "model", i.e. the approximation of reality, is flexible, the overarching aim is to gain a deeper understanding of processes driving the system's behavior often through simulation and analysis of unobserved scenarios. Here, the biomedical reality is described by mathematical formulas or systematic processes purposefully crafted to reflect key aspects of reality with inevitable accompanying simplifying assumptions. For example, dimensionality is reduced, dynamic processes are approximated as time-invariant, or biological pathways are reduced to key components[11]. Conceptualizing these assumptions requires a deep understanding of the biomedical processes and modeling goal. These demands are met through interdisciplinary collaboration and validation. In the absence of experimental data, it is still possible to analyze and simulate to expose dynamics emerging from model building blocks[12–14]. These extrapolations beyond the range of validation data are rooted in the confidence in the quality of the approximation of the biomedical reality, i.e. the quality of the knowledge and its implementation.

It is tempting to suggest that knowledge-driven models are inherently interpretable. Yet, the implementation of chains of relationships can formulate complex inverse problems. Subsequently, *post hoc* processing through parameter identifiability and sensitivity analyses is key[15,16]. This can identify previously unknown interactions between system components to generate hypotheses for experimental and clinical validation.

Knowledge-driven modeling has successfully been applied to investigate different aspects of cancer including somatic cancer evolution and treatment. We refer the interested reader to recent review articles[17,18] covering for instance different fractionation schemes for radiotherapy[19,20], the onset and influence of treatment-induced tumor resistance[21], or cancer evolution[22]. A popular application of knowledge-driven models is the simulation of *in silico* trials for hypothesis generation in simulated cohorts[23–25].

## 2.2. Data-driven models extract information from data

A common understanding of data-driven modeling (e.g. - machine learning, deep learning, and classical statistics) is the creation of insight from empirical examples[26]. A performance metric[27,28] is optimized to uncover patterns and relationships between input data and output task. The validity of data-driven models should be studied carefully, in particular the dependency of the results on the chosen performance metric[28]. It is also key to consider the optimization convergence. If this process fails, the model will be uninformative.

Purely data-driven models do not readily leverage the community's understanding of the system under study but instead employ highly parameterized models. The many degrees of freedom allow flexibility to approximate complex and mechanistically unknown relationships, i.e. these models act as "universal function approximators"[29]. New information can be extracted from the data through this structuring but the extensive parameterization may obscure how the decision process is formed. *Post hoc* processing is required to uncover the nature of the approximated relationship through interpretability and explainability analysis[30]. The models' flexibility also makes them vulnerable to overfitting. Appropriately large amounts of training data and stringent data splits for fitting (training) and validation[31] are necessary to mitigate this risk. Data quantity and quality, i.e. its task specificity and ability to cover a variety of relevant scenarios, are equally important.

Generally the application focus differs from that of knowledge-driven models. Generalization beyond the observed data space is often challenging[32]. It is essential to rely on robust training regimes[33] and consider model limitations as performance is compromised in scenarios not (sufficiently) covered by data[32].

In summary, data-driven approaches are powerful tools for knowledge generation. For example, genome-wide association analysis allows for unbiased evaluation of the importance of different pathways in the context of a cancer and treatment response pattern[34,35], given unbiased training data. Further post-processing on the identified significantly associated genes may reveal new insight into the underlying biological network interactions[36].

## 2.3. Identifying similarities and boundaries between knowledge-driven and data-driven modeling

**Table 1** summarizes and contrasts key characteristics of the extremes of purely data- and knowledge-driven modeling yet boundaries between these models remain fluid for many applications. The fundamental steps of data- and knowledge-driven modeling have parallels despite varying terminology: a subset of data is used to construct and calibrate the model, then further data is necessary for validation and refinement. In data-driven modeling, we first formulate the learning task (i.e. identifying features, labels, and loss function), and architecture selection. In knowledge-driven modeling, we start by deriving equations/mathematical rules. Both algorithms are subsequently compared to real-world data to optimize hyperparameters (i.e., structural model implementations) and to learn model parameters for fitting. The same optimization principles apply but the extent to which mechanistic priors are accounted for in the design of the objective function varies. Finally, validation, ideally on independently sourced data, is performed to assess the model's performance.

*Table 1:* **General conceptual differences between knowledge-driven vs. data-driven modeling.** Some aspects here are taken from Baker et al.[8].

| Knowledge-driven modeling | Data-driven modeling |
|---|---|
| The current "knowledge" drives the implementation of an educated guess regarding the studied relationship. | The empirical reality is approximated through a (complex) relationship. |
| Data serves the purpose of validation of the implemented estimate of reality. | Empirical observations dictate the extraction of information. |
| Generate novel hypotheses for causal mechanisms. | Isolate relevant inputs from empirical datasets for a given output. |
| Deductive capability: extrapolation to predictions about behaviors not present in original data | Inductive capability: interpolation of data with limited extrapolation horizon |
| Predict or describe dynamics of the overall system. | Infer dynamics from the overall system while governing equations and parameters are not exactly known |
| Small but specific data set is needed for validation | Large number of parameters (thousands, millions or more), requiring data-intensive training/fitting |
| Limiting factor(s): Quality of assumptions; parameter sensitivity | Limiting factor(s): Quality and quantity of data; model structure such as choice of features (inputs) |

# 3. Facets of Mechanistic Learning

"Mechanistic learning"[37] can take on many facets by shifting the emphasis of the "data" and "knowledge" paradigms upon model design and fitting. While a partition of mechanistic learning into simulation-assisted machine learning, machine-learning-assisted simulation, and a hybrid class for approaches falling between these definitions is intuitive at first[38], it fails to describe the variety of hybrid approaches. We suggest a more abstract classification (**Figure 1**):

- **Sequential** - Knowledge-based and data-driven modeling are applied sequentially building on the preceding results
- **Parallel** - Modeling and learning are considered parallel alternatives to complement each other for the same objective
- **Extrinsic** - High-level *post hoc* combinations
- **Intrinsic** - Biomedical knowledge is built into the learning approach, either in the architecture or the training phase

While most implementations readily fit into one of these four classes, we emphasize that we do not consider the combinations as discrete encapsulated instances. Instead, we view all synergistic combinations on a continuous landscape between the two extremes of purely knowledge- and data-driven models (**Figure 2**).

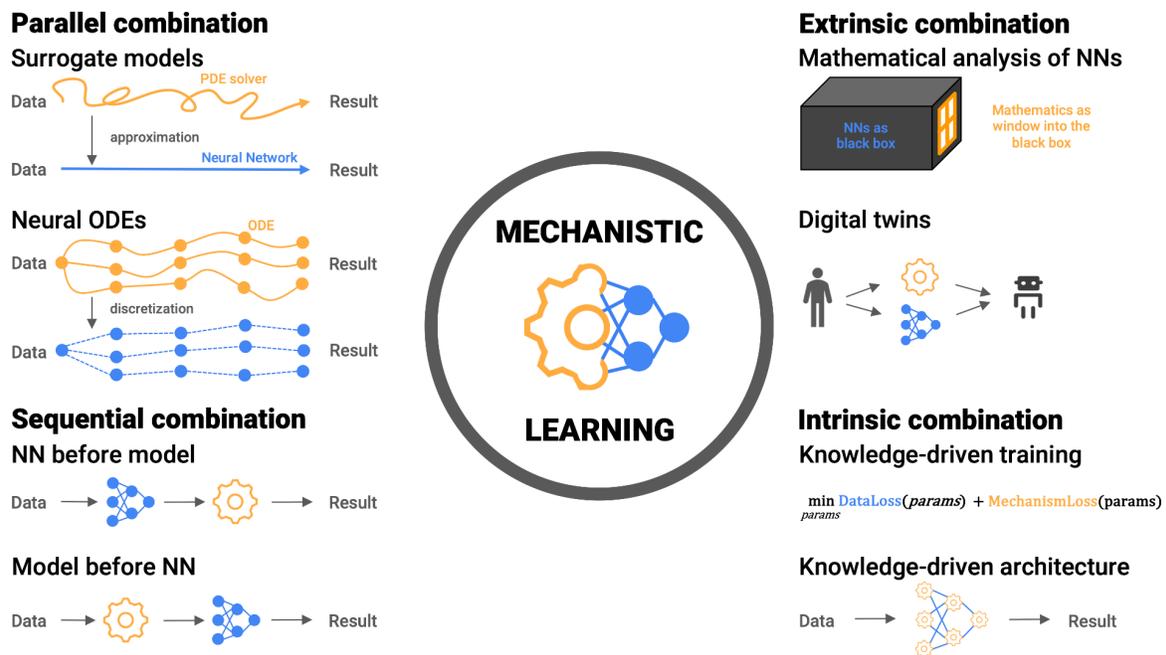

*Fig. 1:* ***Examples of mechanistic learning*** *structured in four combinations:* ***Parallel combinations*** *(top left) with examples of surrogate models and neural ODEs. Data- and knowledge-driven models act as alternatives to complement each other for the same objective.* ***Sequential combinations*** *(bottom left) apply data- and knowledge-driven models in sequence to ease the calibration and validation steps.* ***Extrinsic combinations*** *(top right) combine knowledge-driven and data-driven modeling at a higher level. For example, mathematical analysis of NNs and their results or as complementary tasks for digital twins.* ***Intrinsic combinations*** *(bottom right), like physics- and biology-informed NNs include the knowledge-driven models into the data-driven approaches. Knowledge is included in the architecture of a data-driven model or as regularizer to influence the learned weights.*

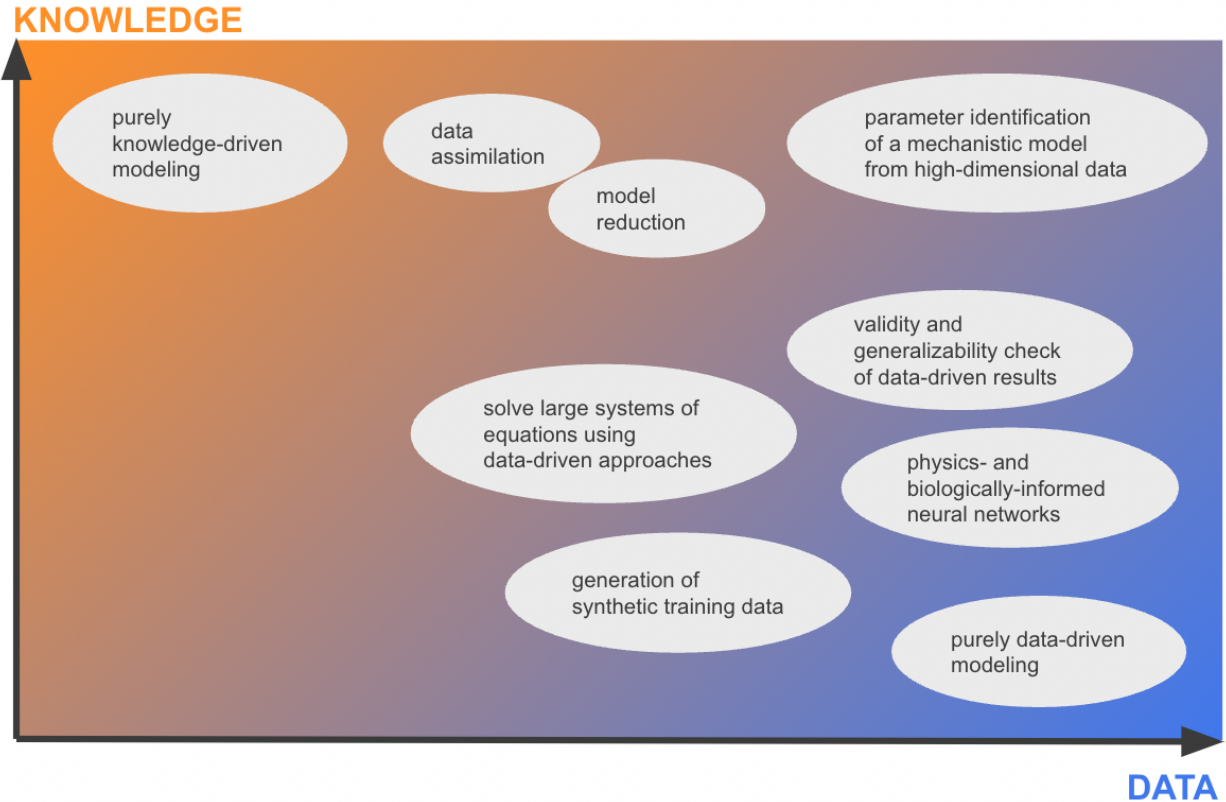

Fig. 2: **The mechanistic learning landscape shows room for the combination of data-driven and knowledge-driven modeling.** *We suggest that purely data-driven or purely knowledge-driven models represent the extremes of a data-knowledge surface with ample room for combinations in different degrees of synergism. Further, in the left-bottom corner with almost no data nor knowledge, any modeling or learning technique is limited.*

## 3.1. Sequential combinations

Sequential approaches harness knowledge and data-driven aspects as sequential and computationally independent tasks by disentangling the parameter/feature estimation and forecasting steps.

### 3.1.1. Domain knowledge to steer data-driven model inputs and architecture choices

In medical science, data availability remains a key challenge[39]. However, there often exists a strong hypothesis regarding the driving features of a specific prediction task. A simple but effective means of improving the performance of data-driven algorithms is a deliberate choice of model architecture, data preprocessing, and model inputs. For example, focusing the input of a deep neural network to disease-relevant subregions of an image boosted classification performance in a data-limited setting[40], and the use of expert-selected features was used to reduce image processing task dimensionality and data requirements of image processing tasks[41]. Similarly important is a deliberate choice of model architecture [42–44]. For instance, while convolution blocks are the staple for computer vision tasks, similar approaches exist for sequential data (e.g. sequence-to-sequence transformers, recurrent NNs, or graph-based models[45,46]). While no mechanistic modeling is conducted *per se* deliberate feature and architecture selection includes additional information. Ultimately, features can also be identified by knowledge-driven modeling[47,48].

### 3.1.2. Mechanistic feature engineering

Feature engineering is the process of designing input features from raw data[49]. This process can be guided by a deeper understanding of the underlying mechanisms, including physical and biochemical laws or causal relationships.

Aspects of a mechanistic model can serve as input features to or outputs from machine learning models. This strategy of "mechanistic feature engineering", was used by Benzekry et al. to predict overall survival in metastatic neuroblastoma patients[50]. First, a mechanistic model of metastatic dissemination and growth was fitted to patient-specific data. Then, a multivariate Cox regression model predicted overall survival from available clinical data with or without patient-specific mechanistic model parameters. They found that including the fitted mechanistic model parameters greatly enhanced the predictive power of the regression.

Similarly, machine learning models can be used to predict the residuals of a mechanistic model's predictions. Kielland et al. utilized this technique to forecast breast cancer treatment outcomes under combination therapy from gene expression data[51]. Initially, a mechanistic model of the molecular mechanisms was calibrated with cell line data to enable patient-specific predictions. Subsequently, various machine learning models were employed to predict the residuals of the mechanistic model from the expression of more than 700 available genes. While the performance of the combined strategy was comparable to using machine learning alone, it offered three advantages. First, the mechanistic model provided a molecular interpretation of treatment response. Additionally, this approach facilitated the discovery of important genes not included in the mechanistic model. Hence, this approach can potentially incorporate emerging biological knowledge and new therapeutics without additional data required for machine learning alone. Note that this sequential strategy facilitates the inclusion of both mechanistically understood features and others that may not be as clear, a common scenario in treatment forecasting.

### 3.1.3. Data-driven estimation of mechanistic model parameters

A common problem in knowledge-driven modeling for longitudinal predictions is parameter identifiability and fitting given limited data and complex systems of equations. The bottleneck lies in the lack of a detailed understanding of the mechanistic relation between input data and desired output, rather than a purely computational limitation. As such, data-driven approaches can be employed to discover correlations within unstructured, high-dimensional data first. These correlations are then harnessed to predict parameters of a mechanistic approach. Importantly, each model is implemented and trained/fitted independently, implying a high-level, yet easily interpretable combination. The sequential combination harnesses the flexibility of data-driven models to handle a variety of data types, and high dimensional inputs to extract information in the form of summarizing parameters that are assigned to a specific contribution upon extrapolation of the knowledge driven contribution. Importantly, the type of data required for such analysis needs to meet the criteria required for knowledge-driven (e.g., longitudinal information) and data-driven (e.g., sufficient sample size) approaches alike. It should be noted that the problem of uncertainty propagation is difficult to address in this truly sequential setting.

In practice, Perez-Aliacar et al.[52] predicted parameters of their mechanistic model of glioblastoma evolution from fluorescent microscopy images. This combination of models has also been suggested in the context of data-driven estimation of pharmacokinetic parameters for drugs[53]. Moreover, data-driven models enable parameter inference by studying parameter dependencies of simulation results through approximate Bayesian computation[54,55] or genetic algorithms[56].

## 3.2 Parallel combinations

Parallel combinations blend advantages of purely data- or knowledge-driven models without changing the anticipated evaluation endpoint.

### 3.2.1 Neural networks as surrogate models

Many phenomena in oncology can be readily formulated using large systems of (partial differential) equations. However, solving large models comes at a high computational cost. Utilizing methods such as model order reduction aids in optimizing the computational efficiency of the solving process. This approach typically demands substantial mathematical expertise and is not suitable for time- or resource-constrained scenarios such as real-world clinical deployment. NNs, as universal function approximators, offer an efficient alternative. In practice, NNs are trained on numerical simulation results and approximate a solution to the system of equations. The inference step of the successfully trained model takes a fraction of the computational resources compared to the full mechanistic model[57,58].

A related concept is the generation of vast amounts of "synthetic" training data[59] based on a small set of "original" data points. While synthetic training data can improve the accuracy of many learning-based systems, care needs to be taken to prevent encoding faulty concepts or misleading biases into the training data that are not present in reality[60,61].

For example, Ezhov et al.[62] introduced a deep learning model performing inverse model inference to obtain the patient-specific spatial distribution of brain tumors from magnetic resonance images, addressing the computational limitations of previous PDE-based spatial tumor growth and response models. A similar brain tumor growth model based on an encoder-decoder architecture trained on 6,000 synthetic tumors generated from a PDE model[63].

### 3.2.2 Neural ODEs — Neural networks as discretized ODEs

The term "neural ODE" describes the notion of viewing neural networks as discretized ODEs or considering ODEs to be neural networks with an infinite amount of layers[64–66]. In that sense, the knowledge-driven approaches using ODEs and the data-driven approach using neural networks are parallel perspectives of the same concept. While not every neural network can be interpreted as discretized ODE and not every question for ODEs can be answered by a discretization to a neural network, neural ODEs can often be a helpful concept to translate between knowledge- and data-driven modeling.

Neural ODEs have already been used for a variety of tasks in oncology ranging from genome-wide regulatory dynamics[67] and breast tumor segmentation in medical images[68] to time-to-event modeling[69].

### 3.2.3 Learning a mechanistic model equation

While oncology research generates vast amounts of data, extracting and consolidating mechanistic understanding from data is a laborious process reliant on human experts. Symbolic regression allows for automated and data-driven discovery of governing laws expressed as algebraic or differential equations. This method finds a symbolic mathematical expression that accurately matches a dataset of label-feature pairs. Two prominent symbolic regression techniques are genetic programming-based optimization[70] and sparse regression[71]. In genetic programming, closed-form expressions are represented as trees and evolved such that trees with high goodness-of-fit are selected for further exploration. In sparse regression strategies, the target expression is assumed to be a linear combination of certain "basis functions", and L1 regularization is used to select and weight a small combination of them.

Despite remarkable success in physics[70], symbolic regression applications in oncology, are still few. In one example, by Brummer et al.[72], sparse regression was employed to estimate a system of ODEs from *in vitro* CAR T-cell glioma therapy data. Compared to knowledge-based models, this data-driven approach offers new insights into the biological dynamics as the model form is not constrained.

Especially in clinical oncology estimating derivatives from high noise and sparse longitudinal measurements remains challenging. Several groups have used variational formulations of ODEs and PDEs in the optimization step without relying on estimating derivatives from noisy and sparse data[73–75]. Bayesian approaches applied to genetic programming have also proven successful in situations where existing non-Bayesian approaches failed[76]. Other promising directions in oncological research are Koopman theory[77] and the universal differential equation framework[78], where neural networks are used to model all or part of a differential equation, facilitating the discovery of governing equations, or parts of them, in cases where data are limited.

## 3.3 Extrinsic combinations

Extrinsic combinations make use of both mechanistic and data-driven approaches to address different aspects of the same problem or to postprocess the output of a data-driven implementation.

### 3.3.1 Digital Twins

Originating from analogies in manufacturing and engineering, the concept of digital twins[79–81] has recently gained interest in the oncology community. A digital twin is an *in silico* patient "twin" that recapitulates important patient characteristics and is used to simulate alternative treatment strategies and forecast disease progression[82]. The computational framework behind the digital twin can be based on a mechanistic, data-driven, or combined set of algorithms. We highlight the potential of combining mechanistic and data-driven modeling as side-by-side tasks, covering different aspects of one unifying digital twin.

Typically, for mechanistic digital twins, a mathematical framework describes the dynamics of tumor size, morphology, composition, and other biomarkers[83]. The data-driven analogy is represented by machine learning algorithms, e.g., k-nearest neighbors but also more advanced architectures, to provide a prediction of the endpoint of interest based on established databases[84,85]. Both knowledge- and data-driven models enable the real-time adaptation of treatment protocols by simulating a range of scenarios. Importantly, harnessing the strengths of each method should be considered for optimal results. For instance, a data-driven prediction task could inform on patient subgrouping and identify likely outcomes, whereas mechanistic modeling would explore personalized treatment alternatives. Generally, digital twins can also serve as "virtual controls" to benchmark the efficacy of the patient's current treatment regimen[86,87].

### 3.3.2 Complementary postprocessing: Mathematical analysis of neural networks and data-driven analysis of mathematical simulations

Neural networks are trained to optimize a performance metric, but performance alone is not driving a model's application in (clinical) practice. Here, quantification of the (un-)certainty of model results, model robustness, as well as interpretability[88] to explain why a neural network arrived at a certain conclusion are equally important. These questions are usually studied under the term explainable AI; for a survey we refer to Roscher et al.[89].

Addressing many of the questions related to neural networks is only possible using mathematical methods, i.e., challenges in the field of NNs are transformed to mathematical conjectures that are subsequently (dis)proven. This approach ensures that the results generated by NNs are mathematically reliable and transparent and thus better suited for clinical implementations.

Further, for the interpretation and validation of simulation results, tools from data-driven modeling can be used to detect patterns in simulations[90]. This approach is already performed in research fields outside the oncology domain[91]. Machine learning and Bayesian statistics can be used for uncertainty quantification which is important for clinical translation[92–94].

## 3.4 Intrinsic combinations

This combination incorporates a mechanistic formulation within a machine learning model either upon training as a contribution to the formulated objective function or *a priori* as a way of choosing the architecture of the neural network. As such, these are densely interconnected combinations.

### 3.4.1 Regularizing the loss function using prior knowledge

Mechanism-informed NNs such as physics-informed neural networks (PINNs)[95,96] use mechanistic regularization upon training, i.e., equation-regularization, by guiding the possible solutions to physically relevant ones. The loss function combines performance loss with a regularization term assessing the deviation from a predefined set of equations. This approach reduces overfitting and ensures physically meaningful predictions. The final NN will not satisfy the equations exactly, but approximate them for the areas where training data is available. PINNs can be valuable for deciding whether an equation can be used to describe data by considering several related equations as regularizers.

Equation-regularization has previously been shown to enhance both the performance and interpretability of data-driven architectures. In the context of oncology, one example includes the modeling of tumor growth dynamics[97]. Ayensa-Jiménez et al[98] used physically-guided NNs with internal variables to model the evolution of glioblastoma as a "go-or-grow" process given constrained resources such as metabolites and oxygen. The model-free nature of their approach allows for the incorporation of data from various boundary conditions and external stimuli, resulting in accurate tumor progression predictions even under different oxygenation conditions.

### 3.4.2 Incorporating knowledge into the machine learning model architecture

Rather than optimizing a network architecture through regularization, biology-informed neural networks constrain the model architecture to biological priors from the start. Typically in the context of network analysis, biological priors such as known interactions between genes and/or transcription factors are translated to nodes and edges in a graph[99,100]. The network is constrained to an established connectivity profile which greatly reduces the model complexity compared to a fully connected network. Similar to transfer learning where a different data-rich scenario is used to pretrain a model prior to refining specific weights on the limited target data, this approach uses expert insight to preset connections and weights. Lagergren et al.[101] proposed biology-informed NNs that learn the nonlinear terms of a governing system, eliminating the need for explicitly specifying the mechanistic form of a PDE as is the case for PINNs. They tested their approach on real-world biological data to uncover previously overlooked mechanisms. Another example is given by Przedborski et al.[102] who used biology-informed NNs to predict patient response to anti-PD-1 immunotherapy and present biomarkers and possible mechanisms of drug resistance. Their model offers insights for optimizing treatment protocols and discovering novel therapeutic targets. Indeed, this approach has found several applications, e.g., for the prediction of prostate cancer[99] and drug discovery[103].

Finally, in the context of generative approaches, differential equations have previously been incorporated into (deep) neural networks through variational autoencoders. While current examples were obtained from medical applications other than oncology[104,105], they represent elegant solutions to allow for dynamic deep learning despite limited data, given careful hyperparameter tuning.

### 3.4.3 Hierarchical Modeling

Hierarchical nonlinear models, also referred to as nonlinear mixed effects models, are a widely used framework to analyze longitudinal measurements on a number of individuals, when interest focuses on individual-specific characteristics[106]. For instance, early in drug development, pharmacokinetics studies are carried out to gain insights into within-subject pharmacokinetics processes of absorption, distribution, and elimination[107]. Typically, a parametric nonlinear model describing drug concentration change over time (individual-level model) is coupled with a linear model describing the relation between pharmacokinetic parameters and individual features (population-level model). One of the simplest population-level models is the random intercept model, which models individual parameter values as normally distributed around a typical value. This enables information sharing through each individual's contribution to determine the typical value, while simultaneously allowing individual parameters that match the observed measurements. Moreover, in contrast to the sequential approach (section 3.1.3), hierarchical models allow for the propagation of uncertainty between the individual-level and population-level models. Applications in oncology range from tumor growth[108] to mutational dynamics in circulating tumor DNA[109] or metastatic dissemination[110].

Interestingly, hierarchical models have the potential to benefit from more sophisticated data-driven approaches to integrate high-throughput data, such as omics or imaging[7]. This can be done by changing the linear covariate model with more complex machine learning algorithms able to capture complex relations between the parameters of the individual-level model and the high dimensional covariates[111,112], and/or by using Bayesian inference[113].

# 4. Conclusion and perspective

Recently, machine and deep learning have become ubiquitous given their indisputable potential to learn from data[114]. However, it is evident that medical applications, especially in oncology, are currently constrained by the extent and diversity of available data. Moreover, clinical translation involves high-stakes decisions that need to be backed up by evidence. The oncology field must address the critical challenges of limited data availability, model transparency, and complex input data. To overcome these bottlenecks, we need data-efficient, comprehensible and robust solutions. Despite the growing interest in mechanistic mathematical modeling for medical applications, the success and opportunity of data-driven models must be taken into account.

Consequently, the thoughtful combination of knowledge- and data-driven modeling is a natural next step.

Here, we identified opportunities for synergistic combinations and provided a snapshot of the current state-of-the-art for how such combinations are facilitated for oncology applications. We highlighted similarities in the mathematical foundation and implementation structure of optimization processes and pointed out differences with respect to data requirements and the role of knowledge and data in these approaches. It is important to structure the growing landscape of models at the interface of data- and knowledge-driven implementations. We hence propose systemizing combinations in four general categories: sequential, parallel, intrinsic, and extrinsic combinations. While sequential and parallel combinations are intuitive and easily implemented, intrinsic and extrinsic combinations incorporate a stronger degree of interlacing that requires a deeper understanding of both data science and mathematical theory. The choice of analysis tool should always keep in mind the quality, size, and type of data and knowledge in light of the underlying research question. An intentional combination of machine learning and mechanistic mathematical modeling can then leverage the strengths of both approaches to tackle complex problems, gain deeper insights, and develop more accurate and robust solutions.

Finally, we hope to motivate a more active exchange between machine learning and mechanistic mathematical modeling researchers given the many parallels in terms of methodologies and evaluation endpoints, and the powerful results produced by mechanistic learning.

# Acknowledgments

We thank Alexander Zeilmann and Saskia Haupt for many fruitful discussions and helpful contributions without which this manuscript would not have been possible. The collaboration that led to the design of this manuscript was fostered during the 2023 Banff International Research Station (BIRS) Workshop on Computational Modelling of Cancer Biology and Treatments (23w5007) initiated by Prof. M. Craig and Dr. A. Jenner.

# Funding statement

JM was supported by NSF 1735095 - NRT: Interdisciplinary Training in Complex Networks and Systems. PM was supported in part by Cancer Moonshot funds from the National Cancer Institute, Leidos Biomedical Research Subcontract 21X126F, and by an Indiana University Luddy Faculty Fellowship. AKL's work was funded by the research centres BigInsight (Norges Forskningsråd project number 237718) and Integreat (Norges Forskningsråd project number 332645). SB was supported by the Botnar Research Center for Child Health Postdoctoral Excellence Programme (#PEP-2021-1008).

## Author Contributions

JM and SB conceived the study, wrote the initial version, and prepared the figures. PM, CJ, and AKL wrote several sections in the text. All authors read and approved the final manuscript.